\documentclass[10pt,twocolumn,amssymb,amsmath,aps,pra]{revtex4-1}
\usepackage{graphicx}

\begin{document}

\title{Mass-polariton theory of sharing the total angular momentum\\ of light between the field and matter}
\date{September 11, 2018}
\author{Mikko Partanen}
\author{Jukka Tulkki}
\affiliation{Engineered Nanosystems group, School of Science, Aalto University, P.O. Box 11000, 00076 Aalto, Finland}

\begin{abstract}
Light propagating in a nondispersive medium is accompanied by a mass density wave (MDW)
of atoms set in motion by the optical force of the field itself
[Phys.~Rev.~A 95, 063850 (2017)]. This recent result is in strong contrast
with the approximation of fixed atoms, which assumes that atoms
are fixed to their equilibrium positions when light propagates in a medium
and which is deeply rooted in the conventional
electrodynamics of continuous media.
In many photonic
materials, the atoms carry the majority of the total momentum of light and their motion
also gives rise to net transfer of medium mass with a light pulse.
In this work, we use optoelastic continuum dynamics
combining the optical force field, elasticity theory, and Newtonian mechanics
to analyze the angular momentum carried by the MDW.
Our calculations are based on classical physics,
but by dividing the numerically calculated angular momenta
of Laguerre-Gaussian (LG) pulses
with the photon number, we can also study the single-quantum values.
We show that
accounting for the MDW in the analysis of the angular momentum
gives for the field's share of the total angular momentum of light a quantized value
that is generally
a fraction of $\hbar$. In contrast, the total angular momentum of the
mass-polariton (MP) quasiparticle, which is a coupled state of the field and
the MDW, and also the elementary quantum of light in a medium, is an integer
multiple of $\hbar$. Thus, the angular momentum of
the MP has coupled field and medium
components, which cannot be separately experimentally measured.
This discovery is related to the previous observation that a bare photon
including only the field part cannot propagate in a medium.
The same coupling
is found for orbital and spin angular momentum components.
The physical picture of the angular momentum of light emerging from our theory
is fundamentally more general than earlier theoretical models, in which the total angular
momentum of light is assumed to be carried by the electromagnetic field only
or by an electronic polariton state, which also involves
dipolar electronic oscillations. These models cannot
describe the MDW shift of atoms associated with light.
We simulate the MDW of LG pulses in silicon and
present a schematic experimental setup for measuring the contribution
of the atomic MDW to the total angular momentum of light.
\end{abstract}

\maketitle

\section{Introduction}

Since the pioneering theoretical work of Allen \emph{et al.} \cite{Allen1992b},
there has been rapid progress in both theoretical and
experimental studies of the angular momentum of light
\cite{Mair2001,Grier2003,Devlin2017,Bozinovic2013,Shao2018,Satoh2017,Du2017,MolinaTerriza2007,Gibson2004,He1995,Friese1998,ONeil2002,CarcesChavez2003,
Beijersbergen1994,Sueda2004,Kotlyar2005,Terhalle2011}.
In the groundbreaking works, the angular momentum of light has been
split into orbital angular momentum (OAM) and spin angular momentum (SAM) \cite{Yao2011,Piccirillo2013,Andrews2013}.
The OAM is related to the helical phasefronts of optical vortex beams
and it is described by the vortex topological charge $l\in\{0,\pm1,\pm2,...\}$,
and the SAM is related to the circular polarization of the wave and it is
described by the polarization helicity $\sigma\in[-1,1]$ \cite{Bliokh2015b,Alpeggiani2018}.

The photon angular momentum has been considered both in vacuum and in
various photonic materials also in the near-field
regime \cite{Lin2013,Bliokh2017a,Bliokh2014}. Advances have been made also in
the understanding of the topological and phase properties of light
\cite{Luo2017,Bhandari1997,Bomzon2002,Bliokh2008a,Bliokh2008b,Bliokh2015b,Ling2017,Ling2015}.
Interestingly, the recently developed mass-polariton (MP) theory of light \cite{Partanen2017c,Partanen2017e}
shows that, in many photonic materials, the majority of the momentum
of light is carried by the medium atoms.
This theory questions the conventional approximation
of fixed atomic positions in the description of propagation of light in a medium.
Thus, a question arises: what happens
in a nondispersive dielectric when the total angular momentum of light
is shared between the electromagnetic field and the medium atoms moving under the
influence of the optical force field?

The MP theory of light shows that a light pulse propagating in
a medium drives forwards an atomic mass density wave
(MDW) \cite{Partanen2017c,Partanen2017e}. The existence of
the MDW that propagates with a light pulse
follows directly from
the classical optoelastic continuum dynamics (OCD),
which combines the well-known optical force density
and the elasticity theory with the Newtonian dynamics of the medium.
In the single-photon picture, the coupling of the electromagnetic
field to the atomic MDW gives rise to MP
quasiparticles, which are covariant coupled states of the field
and matter.

The MP quasiparticles carry a total momentum
of the Minkowski form $p_\mathrm{MP}=n\hbar\omega/c$, where
$n$ is the refractive index of the nondispersive medium,
$\hbar$ is the reduced Planck constant, $\omega$ is the angular
frequency, and $c$ is the speed of light in vacuum \cite{Partanen2017e}.
The total MP momentum is split between the electromagnetic field
and the MDW so that the share of the field corresponds to
the Abraham momentum $p_\mathrm{field}=\hbar\omega/(nc)$
and the MDW carries the difference of the Minkowski and Abraham momenta
$p_\mathrm{MDW}=p_\mathrm{MP}-p_\mathrm{field}$.

In this work, we show that the MP theory of light
can and must be used to describe the angular momentum
of light in nondispersive media.
We will also show that sharing of the angular momentum
between the field and matter is related to fundamental
quantum properties of light.
A single quantum of circularly polarized light is well-known
to carry an angular momentum $\hbar$. Here, we will
show that, e.g., in the case of silicon and assuming
a wavelength of $\lambda_0=1550$ nm, it is shared between
the field ($0.083\hbar$) and the MDW ($0.917\hbar$).
Since both these partial angular momenta
are fractions of $\hbar$, also a fundamental question
arises regarding the quantization of light in a medium.
Our results strongly suggest that only the quantum of the coupled
state of the electromagnetic field and the medium has
a real physical meaning.
We also present a schematic experimental setup for the measurement of the
azimuthal atomic displacements related to the angular momentum transfer
of the MDW in optical fibers.

This work is organized as follows: Section \ref{sec:theory} describes
the theoretical foundations of the description of angular momentum in the MP theory of light.
Section \ref{sec:fields} presents the Laguerre-Gaussian (LG) mode
pulses that are known to carry angular momentum.
The OCD simulations of the propagation of selected LG mode light pulses
in a medium are presented in Sec.~\ref{sec:simulations}. In Sec.~\ref{sec:microparticle},
we discuss whether the atomic MDW has an effect in the analysis of previous
microparticle rotation experiments
\cite{He1995,Friese1996,Simpson1997,Friese1998,ONeil2002,CarcesChavez2003,Adachi2007}.
A schematic plan of the fiber rotation experiment for the experimental
verification of the azimuthal atomic displacement of the MDW
is presented in Sec.~\ref{sec:experimental}.
Finally, conclusions are drawn in Sec.~\ref{sec:conclusions}.

\section{\label{sec:theory}Angular momentum in the mass-polariton theory}

\subsection{Angular momentum of the electromagnetic field}

The fundamental expression of the total angular momentum
of the electromagnetic field in vacuum is conventionally given by
\cite{Jackson1999,Landau1984,Landau1989,Andrews2013,Leader2014,Piccirillo2013,Berry2009,Lenstra1982,vanEnk1992,vanEnk1994,Nienhuis1993,Barnett1994,Barnett2002,Jauregui2005,Calvo2006,Hacyan2006,LiCF2009,Aiello2009,Aiello2010,BialynickiBirula2011,Barnett2010c,Stewart2011,Cameron2015}
\begin{equation}
 \mathbf{J}_\mathrm{field}
 =\int\mathbf{r}\times\Big(\frac{\mathbf{E}\times\mathbf{H}}{c^2}\Big)d^3r,
\label{eq:Jfield}
\end{equation}
where $\mathbf{E}\times\mathbf{H}/c^2=\mathbf{g}_\mathrm{field}$
is the linear momentum density of the electromagnetic field \cite{Jackson1999,Landau1984},
$N_\mathrm{ph}$ is the photon number, and $\hat{\mathbf{z}}$ is the unit vector
in the direction of propagation.
In the MP theory of light, Eq.~\eqref{eq:Jfield}
describes only the electromagnetic field's share of the total angular momentum of light.
This result is independent of the conventional separation
of the angular momentum of the electromagnetic field into SAM and OAM
and into the external and internal parts, which are briefly discussed in
Appendix \ref{apx:separations}.

Note that, in previous works neglecting the atomic MDW \cite{Bliokh2017a,Philbin2013},
the correct total angular
momentum of light in a medium has been obtained
by assuming the Minkowski angular momentum density
$\mathbf{r}\times(\mathbf{D}\times\mathbf{B})$ for the field.
The assumption that the Minkowski angular momentum density
is carried by the pure electromagnetic field
or by the electronic polariton state involving
oscillations of electrons around fixed nuclei
cannot explain the MDW associated with light in a medium
as described by
the MP theory of light \cite{Partanen2017c,Partanen2017e}.

\subsection{Beyond the approximation of fixed atoms}

The approximation of fixed atoms is conventionally
used in the electrodynamics of continuous media \cite{Landau1984,Jackson1999}.
In this approximation, the atomic nuclei are fixed to
their equilibrium positions and they respond to the electromagnetic
field only through polarization. Conventionally, the atoms
have been assumed to be bound to their positions
(or relative positions in the case of moving media)
as their total mass energy is extremely
large compared to the energy scale
of optical fields. Consequently, during the short interaction time
of the medium atoms and the propagating optical field,
the optical force can move atoms only by an exceedingly small amount.
Traditionally, this small atomic movement has been considered
to be totally negligible.

This approximation would be justified
if the polariton state of light would include only
electronic oscillations as is conventionally assumed,
for instance, in the case of exciton polaritons
and in the simple Lorentz oscillator model
of dielectric permittivity.
If we only account for the electronic polariton state,
the center of masses of the negative electron density and the
positive nucleon density move in opposite directions
and the total center of mass of the atom stands still.
Both these mass shifts are anyway, for a free plane-wave field
in a nondispersive medium, considered to be
parallel to the electric field vector, i.e., orthogonal to the
wave vector, and thus, cannot contribute to the MDW,
which propagates in the direction of the Poynting vector.

Recently, the approximation of fixed atoms has been
questioned by the results of the MP theory of light and the related
computer simulations based on the OCD model \cite{Partanen2017c,Partanen2017e}.
Coupling the conventional electrodynamics and elasticity theories of continuous
media, the OCD model shows that the collective small motion
of atoms forms an atomic MDW. Since coupled systems have been under detailed studies
in many fields of physics for decades, it is surprising that the coupled dynamics of
the electromagnetic field and the atomic MDW has not
been studied in detail already much earlier.

We note here an early paper by Poynting
\cite{Poynting1911}, where the existence of ``small longitudinal
material waves accompanying light waves" is foretold. Thus, he considered essentially
the same phenomenon as the MDWs in the MP theory of light. Poynting made
his calculations in the most simple case of an electromagnetic plane wave.
He also correctly calculated the very small kinetic energy associated to the MDWs.
However, he did not calculate the mass energy, which is transferred with
these waves. Therefore, the relation of these waves to the covariance
principle of the special theory of relativity and to the conservation
of the center of energy velocity of an isolated system was not recovered.
In addition, he did not study whether these waves could carry angular momentum.

\subsection{Newton's equation of motion}

In the OCD model, the coupling between the field and matter
is described by Newton's equation of motion.
As the atomic velocities are nonrelativistic,
Newton's equation of motion for the mass density
of the medium $\rho_\mathrm{a}(\mathbf{r},t)$ is given by
\begin{equation}
 \rho_\mathrm{a}(\mathbf{r},t)\frac{d^2\mathbf{r}_\mathrm{a}(\mathbf{r},t)}{dt^2}=\mathbf{f}_\mathrm{opt}(\mathbf{r},t)+\mathbf{f}_\mathrm{el}(\mathbf{r},t),
 \label{eq:mediumnewton}
\end{equation}
where $\mathbf{r}_\mathrm{a}(\mathbf{r},t)$ is the position- and time-dependent atomic
displacement field of the medium, $\mathbf{f}_\mathrm{opt}(\mathbf{r},t)$
is the optical force density experienced by atoms, given in Eq.~\eqref{eq:opticalforcedensity} below, and
$\mathbf{f}_\mathrm{el}(\mathbf{r},t)$ is the elastic force
density between atoms that are displaced from
their initial equilibrium positions by the optical force density.
The elastic force density for anisotropic cubic crystals, such as silicon,
is given, e.g., in Ref.~\cite{Kittel2005}.

For the optical force density, we use the expression
that is well-known in previous literature \cite{Landau1984,Jackson1999}.
It is given for a dielectric medium by \cite{Milonni2010}
\begin{equation}
 \mathbf{f}_\mathrm{opt}(\mathbf{r},t)=-\frac{\varepsilon_0}{2}\mathbf{E}^2\nabla n^2+\frac{n^2-1}{c^2}\frac{\partial}{\partial t}\mathbf{E}\times\mathbf{H}.
 \label{eq:opticalforcedensity}
\end{equation}
Note that the OCD model enables the use of any physically meaninful
form of the optical force density in solving the dynamical
equations \cite{Milonni2010}. However, the different terms cannot be arbitrarily
modified without breaking the momentum conservation
and the relativistic covariance condition \cite{Partanen2017c,Partanen2017e}.

The optical force density in Eq.~\eqref{eq:opticalforcedensity} leads to the transfer of
a part of the total momentum of light by the MDW \cite{Partanen2017c,Partanen2017e}.
In this work, we show that the optical force density in
Eq.~\eqref{eq:opticalforcedensity} is
also responsible for the description of the optical torque
in the medium and the related transfer of a substantial
part of the total angular momentum of light by the MDW.
In calculating the optoelastic force field,
we have neglected the extremely small damping of the
electromagnetic field due to
the transfer of field energy to the kinetic and elastic energies
of the medium. The accuracy of this
approximation is estimated in Ref.~\cite{Partanen2017c}.

\subsection{Angular momentum of the mass density wave}

The angular momentum density of the MDW can be written according to
classical mechanics as
\begin{equation}
 \mathbf{J}_\mathrm{MDW}=\int\mathbf{r}\times\rho_\mathrm{a}\mathbf{v}_\mathrm{a}d^3r,
\label{eq:Jmdw}
\end{equation}
where $\rho_\mathrm{a}\mathbf{v}_\mathrm{a}=\mathbf{g}_\mathrm{MDW}$
is the linear momentum density of the MDW \cite{Partanen2017c}.
This classical angular momentum of the MDW follows purely from the motion of atoms in the MDW
driven by the optical force density. Whether there exist any experimentally feasible ways
to separate this total angular momentum of the MDW into OAM
and SAM parts will be discussed below and is partly left as a topic of further work.

The total angular momentum of the coupled MP state of
the field and matter is given by the sum of the field and MDW contributions as
\begin{equation}
 \mathbf{J}_\mathrm{MP}=\int\mathbf{r}\times\Big(\rho_\mathrm{a}\mathbf{v}_\mathrm{a}+\frac{\mathbf{E}\times\mathbf{H}}{c^2}\Big)d^3r,
\label{eq:Jmp}
\end{equation}
where $\rho_\mathrm{a}\mathbf{v}_\mathrm{a}+\mathbf{E}\times\mathbf{H}/c^2=\mathbf{g}_\mathrm{MP}$
is the total linear momentum density of the MP.
As shown by the computer simulations below,
the angular momentum of the MDW is an integral
part of the total angular momentum of light in a medium.

\section{\label{sec:fields}Fields carrying angular momentum}

\subsection{Laguerre-Gaussian pulses}

It is well known that helically phased light carries OAM regardless of the radial
distribution of the fields \cite{Yao2011}. However, it is often useful to express beams
in a complete basis set of orthogonal modes. A convenient choice as the basis set
for beams that carry OAM is provided by the LG modes.

The LG mode function for the mode
$\mathrm{LG}_{pl}$ is given in the limit of constant
beam waist or large Rayleigh range as \cite{Allen1992b,Cerjan2011}
\begin{equation}
 u_{p,l}(r,\phi)=u_0\Big(\frac{\sqrt{2}r}{w_0}\Big)^{|l|} e^{-r^2/w_0^2}e^{il\phi}L_p^{|l|}\Big(\frac{2r^2}{w_0^2}\Big),
\label{eq:LaguerreGaussian}
\end{equation}
where $L_p^l$ are the generalized Laguerre polynomials,
$u_0$ is a normalization constant that also depends on the values of $p$ and $l$,
and $w_0$ is the waist radius \cite{Allen1992b}.
The waist radius is a distance where the intensity of the beam
has dropped to $1/e^2$ of its on-axis value.

\subsection{Linear polarization}

The electric and magnetic fields of a LG
beam with linear polarization along the $x$ direction
are written in the paraxial approximation for a given wavenumber $k$
as presented, e.g., in Refs.~\cite{Allen2011,Allen1992b,Haus1984}.
Here, we study a light pulse with a Gaussian distribution for $k$ as described by
the Gaussian function $u(k)=e^{-[(k-nk_0)/(n\Delta k_0)]^2/2}/(\sqrt{2\pi}\,n\Delta k_0)$,
where $k_0=\omega_0/c$ is the wavenumber in
vacuum for central frequency $\omega_0$, and $\Delta k_0$
is the standard deviation of the wavenumber in vacuum.
Therefore, we write the electric and magnetic fields of
the $\mathrm{LG}_{pl}$ pulse as
\begin{align}
 &\mathbf{E}_{p,l}(\mathbf{r},t)\nonumber\\
 &=\mathrm{Re}\Big[\int_{-\infty}^\infty i\omega(k)\Big(u_{p,l}\hat{\mathbf{x}}+\frac{i\partial u_{p,l}}{k\partial x}\hat{\mathbf{z}}\Big)u(k)e^{i[kz-\omega(k) t]}dk\Big],
 \label{eq:efield1}
\end{align}
\begin{align}
 &\mathbf{H}_{p,l}(\mathbf{r},t)\nonumber\\
 &=\mathrm{Re}\Big[\int_{-\infty}^\infty\frac{ik}{\mu_0}\Big(u_{p,l}\hat{\mathbf{y}}+\frac{i\partial u_{p,l}}{k\partial y}\hat{\mathbf{z}}\Big)u(k)e^{i[kz-\omega(k) t]}dk\Big]
 \label{eq:hfield1},
\end{align}
where $\omega(k)=ck/n$ is the dispersion relation of a nondispersive medium.

The wavenumber and its standard deviation in the medium are
given in terms of the vacuum quantities by $k_\mathrm{0,med}=nk_0$
and $\Delta k_\mathrm{0,med}=n\Delta k_0$. The standard deviations
of the $x$ and $y$ components of the wavenumber are given by
$\Delta k_x=\Delta k_y=\sqrt{2}/w_0$.
The standard deviations of the spatial dimensions of the pulse
energy density are $\Delta z=1/(\sqrt{2}\Delta k_\mathrm{0,med})$
and $\Delta x=\Delta y=1/(\sqrt{2}\Delta k_x)=w_0/2$.
The corresponding standard deviation in
time is then $\Delta t=n\Delta z/c=1/(\sqrt{2}\Delta k_\mathrm{0}c)$ and the full width
at half maximum is $\Delta t_\mathrm{FWHM}=2\sqrt{2\ln 2}\Delta t$.

In the monochromatic field limit, the fields in Eqs.~\eqref{eq:efield1} and \eqref{eq:hfield1}
and the resulting Poynting vector can be approximated further as explained
in Appendix \ref{apx:approximation}. This approximation allows us to perform
the integrations in  Eqs.~\eqref{eq:efield1} and \eqref{eq:hfield1} analytically,
which reduces the computational power needed in the three-dimensional simulations.
As shown in Ref.~\cite{Partanen2017c}, we can also approximate the actual instantaneous Poynting
vector of a light pulse with its time-average over the harmonic cycle without
losing accuracy in the calculation of the total quantities, such as the linear and angular momenta
and the transferred mass of the light pulse.

\subsection{Circular polarization}

As defined from the point of view of the source,
the electric and magnetic fields of a right
circularly polarized $\mathrm{LG}_{pl}$ pulse are given
in the paraxial approximation by \cite{Kenyon2008,Allen2011}
\begin{align}
 &\mathbf{E}_{p,l}(\mathbf{r},t)\nonumber\\
 &=\frac{1}{\sqrt{2}}\mathrm{Re}\Big[\int_{-\infty}^\infty\! i\omega(k)\Big(u_{p,l}\hat{\mathbf{x}}+\frac{i\partial u_{p,l}}{k\partial x}\hat{\mathbf{z}}\Big)u(k)e^{i[kz-\omega(k) t]}dk\nonumber\\
 &\hspace{0.4cm}+\int_{-\infty}^\infty\! i\omega(k)\Big(u_{p,l}\hat{\mathbf{y}}+\frac{i\partial u_{p,l}}{k\partial y}\hat{\mathbf{z}}\Big)u(k)e^{i[kz-\omega(k) t+\pi/2]}dk\Big]
\label{eq:efield2},
\end{align}
\begin{align}
 &\mathbf{H}_{p,l}(\mathbf{r},t)\nonumber\\
 &=\frac{1}{\sqrt{2}}\mathrm{Re}\Big[\int_{-\infty}^\infty\frac{ik}{\mu_0}\Big(u_{p,l}\hat{\mathbf{y}}+\frac{i\partial u_{p,l}}{k\partial y}\hat{\mathbf{z}}\Big)u(k)e^{i[kz-\omega(k) t]}dk\nonumber\\
 &\hspace{0.4cm}-\int_{-\infty}^\infty\frac{ik}{\mu_0}\Big(u_{p,l}\hat{\mathbf{x}}+\frac{i\partial u_{p,l}}{k\partial x}\hat{\mathbf{z}}\Big)u(k)e^{i[kz-\omega(k) t+\pi/2]}dk\Big]
\label{eq:hfield2}.
\end{align}
These fields can be seen as superpositions of two linearly polarized fields,
whose transverse components are orthogonal to each other, and which
have a phase difference of $\pi/2$. The first terms in Eqs.~\eqref{eq:efield2} and \eqref{eq:hfield2}
correspond to the linearly polarized fields in Eqs.~\eqref{eq:efield1} and \eqref{eq:hfield1}.

As in the case of linear polarization above,
in the monochromatic field limit, the fields in Eqs.~\eqref{eq:efield2} and \eqref{eq:hfield2}
and the resulting Poynting vector can be approximated further as explained
in Appendix \ref{apx:approximation}.

\begin{figure*}
\includegraphics[width=0.98\textwidth]{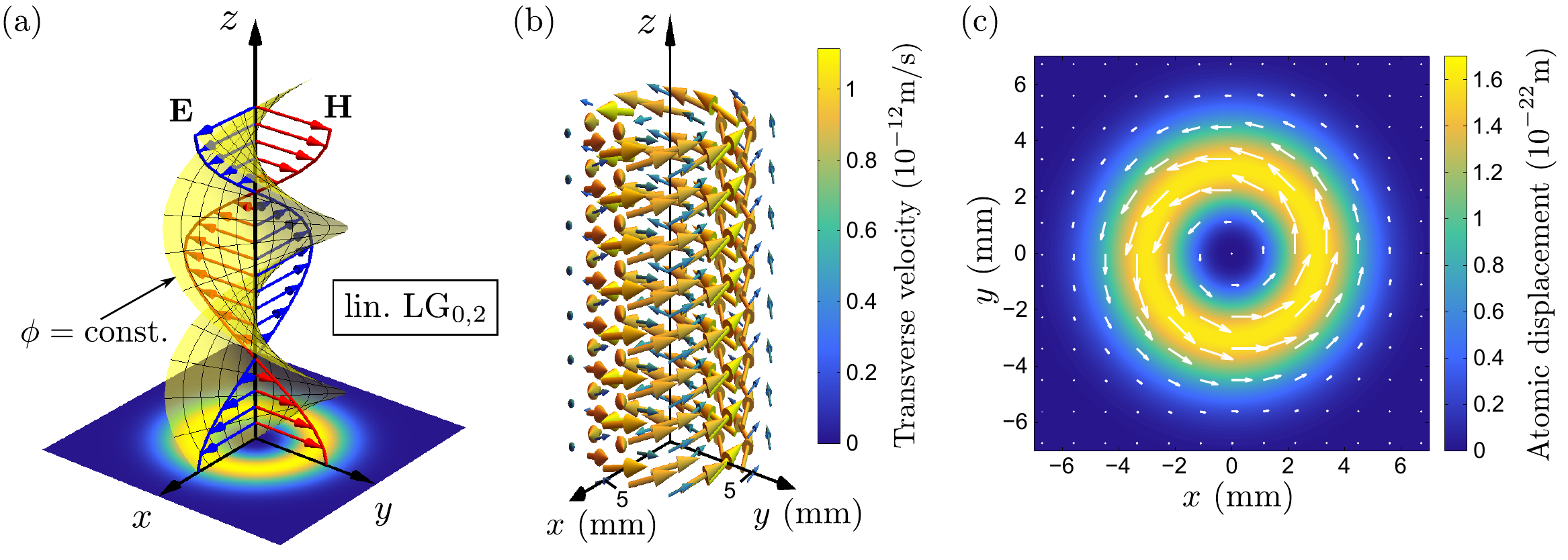}
\caption{\label{fig:oam}
Simulations for a  linearly polarized LG$_{0,2}$
pulse that carries OAM in silicon.
(a) The electric field polarization lies in the vertical
plane $y=0$ as the polarization is linear.
The opaque surface shows the helical phasefront of the pulse.
The density plot in the plane $z=0$ shows the
time-averaged field intensity.
(b) The vector arrows present the time-averaged
atomic velocities in the MDW driven by the field.
In the $z$ direction, the plot region corresponds to one harmonic cycle
in the middle of the pulse. The $z$ component
of the atomic velocities dominates and it has been scaled down
using a factor of $10^{-5}$ to make the spiraling of the
atomic velocities around the optical axis visible.
The colors correspond to the magnitude of the
transverse component of the atomic velocities.
(c) Representation of the azimuthal atomic displacement
due to the transfer of OAM with the MDW of atoms.
The vector arrows show the direction and the color bar shows
the magnitude. The atomic displacement is represented
at the instance of time just after the pulse has gone
and elastic forces have not had time enought to relax
the strain field related to these atomic displacements.}
\end{figure*}

\section{\label{sec:simulations}Angular momentum simulations}

Next we simulate the propagation of LG light pulses
in silicon. We simulate a pulse, which has a total electromagnetic energy of $U_0=5$ mJ
and a central vacuum wavelength of $\lambda_0=1550$ nm. These values correspond
to the central angular frequency of $\omega_0=2\pi c/\lambda_0=1.215\times 10^{15}$ s$^{-1}$ and the
photon number of $N_\mathrm{ph}=U_0/\hbar\omega_0=3.901\times 10^{16}$.
The phase and group refractive indices of silicon are given by
$n_\mathrm{p}=3.4757$ and 
$n_\mathrm{g}=3.5997$ for $\lambda_0=1550$ nm \cite{Li1980}.
Since the dispersion is not very large, for simplicity,
in this work we neglect the dispersion and use the phase refractive index only.
However, the dispersion could be accounted for in the MP theory
of light as shown in Ref.~\cite{Partanen2017c}.

The lateral width of the LG mode in Eq.~\eqref{eq:LaguerreGaussian}
is determined by $\Delta k_x=\Delta k_y=\sqrt{2}/w_0=10^{-4}k_0$,
which correspond to the waist radius of $w_0\approx 3.5$ mm.
The longitudinal pulse width is determined by
the spectral width $\Delta\omega/\omega_0=\Delta k_0/k_0=10^{-5}$,
which corresponds to $\Delta z\approx 5.0$ mm and
$\Delta t_\mathrm{FWHM}\approx 140$ ps.
The normalization constant $u_0$ in Eq.~\eqref{eq:LaguerreGaussian}
becomes fixed by requiring that the total electromagnetic energy of the pulse is $U_0$
as defined above.

In the simulations, we also use the mass density of silicon, which
is $\rho_0=2329$ kg/m$^3$ \cite{Lide2004}, and the elastic
constants in the direction of the (100) plane, which are
$C_{11}=165.7$ GPa, $C_{12}=63.9$ GPa, and $C_{44}=79.6$ GPa
\cite{Hopcroft2010}. These elastic constants correspond to the
bulk modulus of $B=(C_{11}+2C_{12})/3=97.8$ GPa and the shear modulus of $G=C_{44}=79.6$ GPa.

\subsection{Simulations for orbital angular momentum}

First, we consider the OAM related to the MDW of moving atoms.
As an example, we use a linearly polarized
$\mathrm{LG}_{0,2}$ pulse that is well known to carry OAM as $l=2$.
The SAM of this pulse is zero due to linear polarization
for which $\sigma=0$.

Figure \ref{fig:oam}(a) illustrates the linearly polarized $\mathrm{LG}_{0,2}$ mode.
The vector arrows in Fig.~\ref{fig:oam}(a)
present the directions of the instantaneous electric (blue)
and magnetic (red) fields. For our linearly polarized field, the electric field vectors lie in the
plane $y=0$ while the magnetic field vectors are located
in the plane $x=0$. The exact spatial distributions of the
fields are, however, more complex and not shown in the figure.
The opaque surface shows the phasefront
of the pulse, which forms a double helix for our pulse with $l=2$.
The density plot in the plane $z=0$ shows
the time-averaged field intensity.
The field intensity has a vortex at $x=y=0$, which
is characteristic for all higher order LG modes.

Figure \ref{fig:oam}(b) shows the time-averaged atomic velocities in the MDW
driven by a linearly polarized $\mathrm{LG}_{0,2}$ pulse. The $z$ component
of the atomic velocities dominates and it has been scaled down
using a factor of $10^{-5}$ to make the transverse velocity components visible.
The atomic velocities are seen to spiral around the optical axis
so that the velocity distribution of atoms has a vortex at $x=y=0$.
The velocity distribution of atoms in the MDW clearly follows the field intensity
in the density plot in Fig.~\ref{fig:oam}(a). This is an expected result
since the optical force density in the second term of
Eq.~\eqref{eq:opticalforcedensity} driving the atomic MDW forwards
is determined by the Poynting vector.

\begin{figure*}
\includegraphics[width=0.98\textwidth]{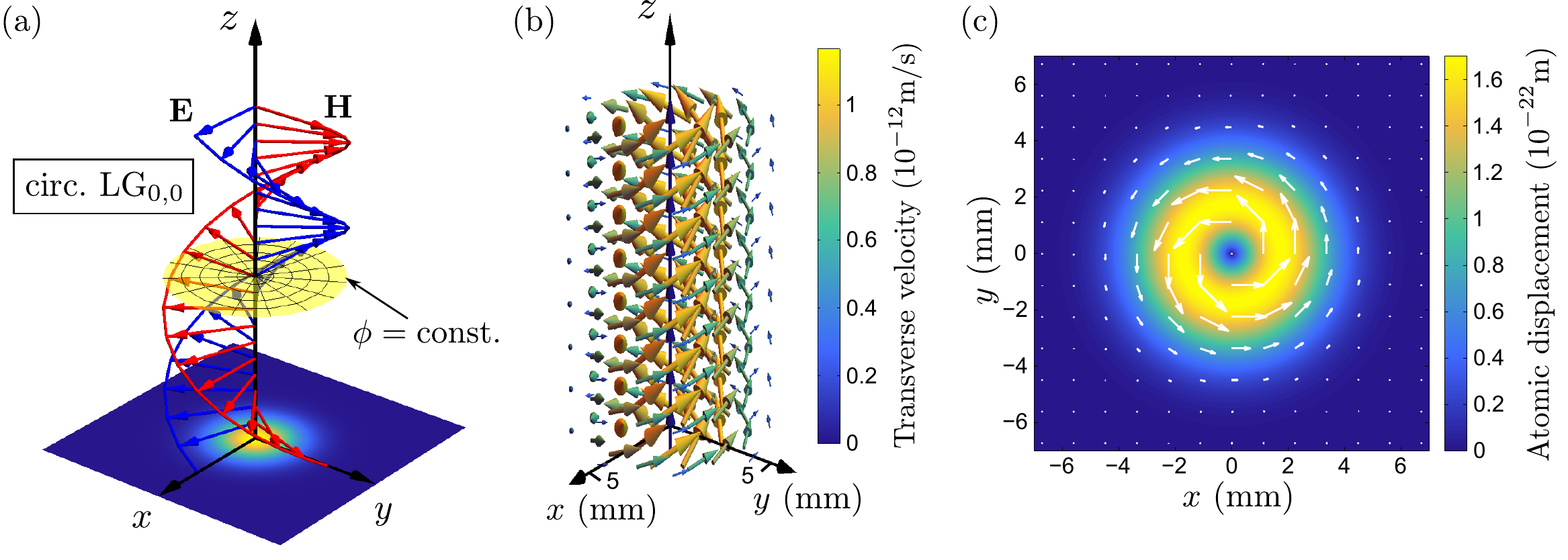}
\caption{\label{fig:sam}
Simulations for a right circularly polarized LG$_{0,0}$ pulse
that carries SAM in silicon.
(a)  The position dependence of the transverse component of the
electric field polarization forms
a helical surface as the polarization is circular.
The opaque surface shows the plane of constant phase.
The density plot in the plane $z=0$ shows
the time-averaged field intensity.
(b) The vector arrows present the time-averaged atomic
velocities in the MDW driven by the field.
In the $z$ direction, the plot region corresponds to one harmonic cycle
in the middle of the pulse.
The $z$ component
of the atomic velocities dominates and it has been scaled down
using a factor of $10^{-5}$ to make the spiraling of the
atomic velocities around the optical axis visible.
The colors correspond to the magnitude of the
transverse component of the atomic velocities.
(c) Representation of the azimuthal atomic displacement
due to the transfer of OAM with the MDW of atoms.
The vector arrows show the direction and the color bar shows
the magnitude. The atomic displacement is represented
at the instance of time just after the pulse has gone
and elastic forces have not had time enought to relax
the strain field related to these atomic displacements.}
\end{figure*}

The atomic velocity distribution of the light pulse obtained
in the simulations can be used with Eq.~\eqref{eq:Jmdw}
to calculate the angular momentum of the MDW.
Within the numerical accuracy of the simulations and the
monochromatic field approximation described in Appendix \ref{apx:approximation},
we obtain $\mathbf{J}_\mathrm{MDW}=7.157\times 10^{16}\hbar\hat{\mathbf{z}}$.
For the angular momentum of the electromagnetic field, given by Eq.~\eqref{eq:Jfield},
we obtain $\mathbf{J}_\mathrm{field}=6.459\times 10^{15}\hbar\hat{\mathbf{z}}$.
Dividing these values with the photon number of the pulse
gives $\mathbf{J}_\mathrm{MDW}/N_\mathrm{ph}=1.834\hbar\hat{\mathbf{z}}$
and $\mathbf{J}_\mathrm{field}/N_\mathrm{ph}=0.166\hbar\hat{\mathbf{z}}$.
The sum of these angular momenta is the total angular momentum of the mass-polariton,
given by $\mathbf{J}_\mathrm{MP}/N_\mathrm{ph}=2\hbar\hat{\mathbf{z}}$.
This equals the expected total angular momentum of the linearly
polarized $\mathrm{LG}_{0,2}$ mode with $l=2$ and $\sigma=0$. Therefore, our results indicate
that the total angular momentum of light in a medium is split between the field
and the MDW in such a way that the MDW carries a substantial part
of the total angular momentum of light.

Figure \ref{fig:oam}(c) shows the transverse component of the
simulated atomic displacement for the linearly polarized
$\mathrm{LG}_{0,2}$ pulse just after the light pulse has gone.
The time-dependent simulations of the atomic displacements and
velocities in a fixed transverse plane are presented as a
video file in the Supplementary Material \cite{supplementaryvideo}.
In the passage of the pulse, the atomic displacements monotonically
increase to their maximum values, whose transverse components are shown in \ref{fig:oam}(c).
In contrast, the average atomic velocity increases in the pulse front, obtains
its maximum value in the middle of the pulse, and decreases to
zero in the tail of the pulse.

\subsection{Simulations for spin angular momentum}

Next, we consider the SAM of the atomic MDW by simulating the propagation of a right
circularly polarized $\mathrm{LG}_{0,0}$ pulse. This pulse carries SAM but no OAM
as $l=0$ and $\sigma=1$.

Figure \ref{fig:sam}(a) illustrates the right circularly polarized
$\mathrm{LG}_{0,0}$ mode.
The vector arrows show the directions of the instantaneous
electric (blue) and magnetic (red) fields. These field vectors
form polarization surfaces that are helical for our circularly polarized field.
The opaque surface shows the phasefront
of the pulse. In contrast to the higher order $\mathrm{LG}_{0,2}$ pulse
in Fig.~\ref{fig:oam}(a), the phasefront is a plane for our
$\mathrm{LG}_{0,0}$ pulse with $l=0$.
The density plot in the transverse plane $z=0$ shows the
cross section of the time-averaged field intensity.
In contrast to the case of the $\mathrm{LG}_{0,2}$
pulse in Fig.~\ref{fig:oam}(a),
the field intensity in Fig.~\ref{fig:sam}(a)
obtains its maximum value at $x=y=0$
for the $\mathrm{LG}_{0,0}$ pulse.
Therefore, there is not any vortex in the field intensity
when $l=0$.

Figure \ref{fig:sam}(b) shows the time-averaged atomic velocities in the MDW
driven by the right circularly polarized $\mathrm{LG}_{0,0}$ pulse.
The atomic velocities are again seen to spiral around the direction
of propagation along the $z$ axis. Again, the $z$ component
of the atomic velocities dominates and it has been scaled down
using a factor of $10^{-5}$ to make the transverse velocity components visible.
Whereas, in Fig.~\ref{fig:oam},
the spiraling of the atomic velocities followed from the higher order nature
of the $\mathrm{LG}_{0,2}$ pulse, here the spiraling
follows purely from the circular polarization.
As in the case of OAM in Fig.~\ref{fig:oam},
the atomic velocities in the MDW driven by the circularly polarized light pulse
carrying only SAM also follow the field intensity in the density plot in Fig.~\ref{fig:sam}(a).
Again, this follows from the fact that the optical force density in the second term of
Eq.~\eqref{eq:opticalforcedensity} driving the atomic MDW forwards
is determined by the Poynting vector.

Using Eq.~\eqref{eq:Jmdw} and the atomic velocity distribution of the MDW obtained
in the simulations, we can again calculate the angular momentum of the MDW.
Within the monochromatic field approximation described in Appendix \ref{apx:approximation}
and the numerical accuracy of the simulations,
we obtain $\mathbf{J}_\mathrm{MDW}=3.578\times 10^{16}\hbar\hat{\mathbf{z}}$.
The corresponding angular momentum of the electromagnetic field, given by Eq.~\eqref{eq:Jfield},
has a value of $\mathbf{J}_\mathrm{field}=3.230\times 10^{15}\hbar\hat{\mathbf{z}}$.
Dividing the angular momenta of the MDW and the electromagnetic field with
the photon number of the pulse gives
$\mathbf{J}_\mathrm{MDW}/N_\mathrm{ph}=0.917\hbar\hat{\mathbf{z}}$
and $\mathbf{J}_\mathrm{field}/N_\mathrm{ph}=0.083\hbar\hat{\mathbf{z}}$.
Summing these angular momenta together then gives total angular momentum of the mass-polariton
as $\mathbf{J}_\mathrm{MP}/N_\mathrm{ph}=\hbar\hat{\mathbf{z}}$,
which is an expected result for the right circularly
polarized $\mathrm{LG}_{0,0}$ mode with $l=0$ and $\sigma=1$. Therefore,
these results give further support for the splitting of the total
angular momentum of light between the field and the MDW.

\begin{figure*}
\includegraphics[width=\textwidth]{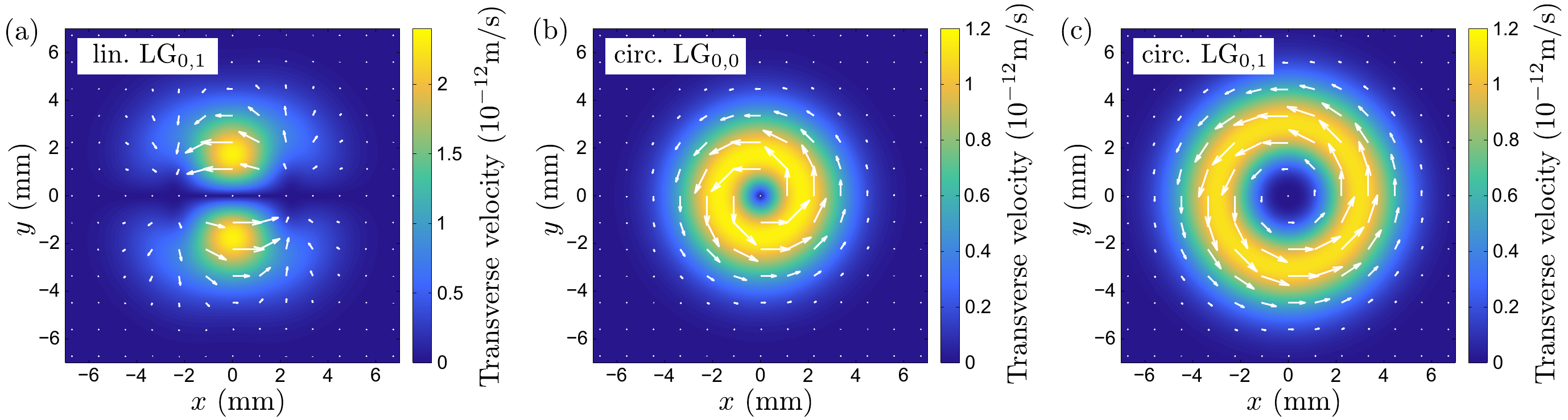}
\caption{\label{fig:instantaneous}
Instantaneous atomic velocity in the MDW. The
position dependence of the transverse component of the
instantaneous atomic velocity of the MDW (a) for the linearly polarized $\mathrm{LG}_{0,1}$ pulse
carrying OAM of $N_\mathrm{ph}\hbar$,
(b) for the circularly polarized $\mathrm{LG}_{0,0}$ pulse carrying SAM of $N_\mathrm{ph}\hbar$, and
(c) for the circularly polarized $\mathrm{LG}_{0,1}$ pulse
carrying both OAM of $N_\mathrm{ph}\hbar$ and SAM of $N_\mathrm{ph}\hbar$.
The transverse atomic velocities are plotted for one instance of time
in the middle of the pulse.}
\end{figure*}

Figure \ref{fig:sam}(c) shows the transverse component of the
simulated atomic displacement for the right circularly
polarized $\mathrm{LG}_{0,0}$
pulse just after the light pulse has gone.
The time-dependent simulations of the atomic displacements and
velocities in a fixed transverse plane are presented as a
video file in the Supplementary Material \cite{supplementaryvideo}.
In the passage of the pulse, the atomic displacements again monotonically
increase to their maximum values, while the average atomic velocity increases in the pulse front, obtains
its maximum value in the middle of the pulse, and decreases to
zero in the tail of the pulse.

\subsection{Differences in the instantaneous mass density waves}

Figure \ref{fig:instantaneous} shows the transverse components of the
instantaneous atomic velocities in the MDWs of the linearly polarized
$\mathrm{LG}_{0,1}$ pulse carrying OAM ($N_\mathrm{ph}\hbar$), the circularly polarized $\mathrm{LG}_{0,0}$
pulse carrying SAM ($N_\mathrm{ph}\hbar$), and the circularly polarized $\mathrm{LG}_{0,1}$ pulse
carrying both OAM ($N_\mathrm{ph}\hbar$) and SAM ($N_\mathrm{ph}\hbar$).
One can directly observe the notable difference that,
in the case of linear polarization in Fig.~\ref{fig:instantaneous}(a)
(see also a video file in the Supplementary Material \cite{supplementaryvideo}), the atomic velocities
vary as a function of the azimuthal angle, whereas in the case of circular polarizations in
Figs.~\ref{fig:instantaneous}(b) and \ref{fig:instantaneous}(c), this azimuthal dependence is missing.
Therefore, we can conclude that the MDW component that has azimuthal
variations is related to the linear polarization while the MDW component
that is approximately constant in the azimuthal direction is related
to the circular polarization. Thus, the MDWs of linearly and circularly
polarized light pulses are in principle separable. However, whether the physical
separation into the OAM and SAM components is possible in the general case
of a pulse carrying both OAM and SAM remains an open question
in the classical regime where one can in principle measure the
transverse velocity distribution and the shift of atoms in the MDW.
However, for a single light quantum in a medium, our results strongly suggest
that the total angular momentum of the quantum, which is
an integer multiple of $\hbar$, cannot be divided in a physically
meaningful way into the OAM and SAM components of either
the MDW or the electromagnetic field.

\section{\label{sec:comparison}Comparison of the OCD and MP quasiparticle model results}

For a general Laguerre-Gaussian pulse with total electromagnetic
energy $U_0=N_\mathrm{ph}\hbar\omega_0$, we can calculate
the numerical values of the angular momenta in Eqs.~\eqref{eq:Jfield},
\eqref{eq:Jmdw}, and \eqref{eq:Jmp}. Within the numerical
accuracy of the simulations, we obtain
\begin{equation}
 \mathbf{J}_\mathrm{field}=\int\mathbf{r}\times\Big(\frac{\mathbf{E}\times\mathbf{H}}{c^2}\Big)d^3r=(l+\sigma)\frac{N_\mathrm{ph}\hbar}{n^2}\hat{\mathbf{z}},
 \label{eq:comparisonJfield}
\end{equation}
\begin{equation}
 \mathbf{J}_\mathrm{MDW}=\int\mathbf{r}\times\rho_\mathrm{a}\mathbf{v}_\mathrm{a}d^3r=\Big(1-\frac{1}{n^2}\Big)(l+\sigma)N_\mathrm{ph}\hbar\hat{\mathbf{z}},
 \label{eq:comparisonJmdw}
\end{equation}
\begin{equation}
 \mathbf{J}_\mathrm{MP}=\int\mathbf{r}\times\Big(\rho_\mathrm{a}\mathbf{v}_\mathrm{a}+\frac{\mathbf{E}\times\mathbf{H}}{c^2}\Big)d^3r=(l+\sigma)N_\mathrm{ph}\hbar\hat{\mathbf{z}},
 \label{eq:comparisonJmp}
\end{equation}
i.e., the numerically calculated values of the integrals are
equal to the right hand side results within 7 digits.
Thus, the right hand side results divided by the photon number
represent the single quantum MP quasiparticle model values
of the corresponding angular momenta.

Equations \eqref{eq:comparisonJfield}--\eqref{eq:comparisonJmp}
indicate that the total angular momentum of light in a nondispersive
medium is split
between the field and the MDW in the same ratio
$\mathbf{J}_\mathrm{MDW}/\mathbf{J}_\mathrm{field}=n^2-1$
as the total linear momentum $p_\mathrm{MDW}/p_\mathrm{field}$
and energy $E_\mathrm{MDW}/E_\mathrm{field}$ shown in
Refs.~\cite{Partanen2017c,Partanen2017e}.
The comparison of Eqs.~\eqref{eq:comparisonJfield}--\eqref{eq:comparisonJmp}
shows that the atomic MDW carries a substantial part (for many photonic materials most)
of the total angular momentum of light in a medium.
In previous works on the angular momentum of light
in a medium \cite{Bliokh2017a,Philbin2013},
which have neglected the atomic MDW,
the correct total angular momentum of light has been obtained
by assuming that the field carries the Minkowski momentum,
which corresponds to the angular momentum density
$\mathbf{r}\times(\mathbf{D}\times\mathbf{B})$.
According to the MP theory of light, this assumption
is not justified \cite{Partanen2017c,Partanen2017e}.

\section{\label{sec:microparticle}Analyzing microparticle rotation experiments}

It is worth to consider whether the MDW has an effect in the previous
microparticle rotation experiments performed to probe the OAM and SAM of light
\cite{He1995,Friese1996,Simpson1997,Friese1998,ONeil2002,CarcesChavez2003,Adachi2007}.
These experiments are based on trapping particles by using optical tweezers,
which rely on the gradient force to confine a dielectric particle near the point
of highest light intensity \cite{Ashkin1986,ONeil2002}. It has been observed that 
the OAM and SAM of light result in the rotation of absorptive
\cite{He1995,Friese1996,Simpson1997} and birefringent
\cite{Friese1998,ONeil2002,CarcesChavez2003} particles.
Therefore, the physical
mechanism of the observed angular momentum transfer is very different compared
to the angular momentum of the MDW. In order to observe the angular momentum
of the MDW, one should aim at observing the rotation of the homogeneous host medium
itself when light propagates through it. This rotation related to the MDW is much smaller
than the rotation of a single microparticle and it is not expected to be
observable in the previous experiments.

\section{\label{sec:experimental}Planning of the fiber rotation experiment}

Next, we study how the azimuthal atomic displacement due to the MDW simulated in
Sec.~\ref{sec:simulations} could be experimentally verified. For this purpose,
we propose a fiber rotation experiment, where one measures the rotation
of an optical fiber due to optical forces of circularly polarized
continuous wave light beams after the fields have been switched on at time $t=0$ s.
These calculations can also inspire experimentalists
to design other kinds of setups for the same purpose
of verifying the azimuthal atomic displacement predicted by the MP theory of light.
In the relatively long time scale used here, the strain fields
in the fiber cross section are assumed to be relaxed by the elastic
forces so that all material element in the fiber cross section have
approximately the same angle of rotation with respect to the fiber axis.
In particular, here we focus on investigating the azimuthal atomic displacement
that should be measured at the fiber surface.

\begin{figure}
\includegraphics[width=\columnwidth]{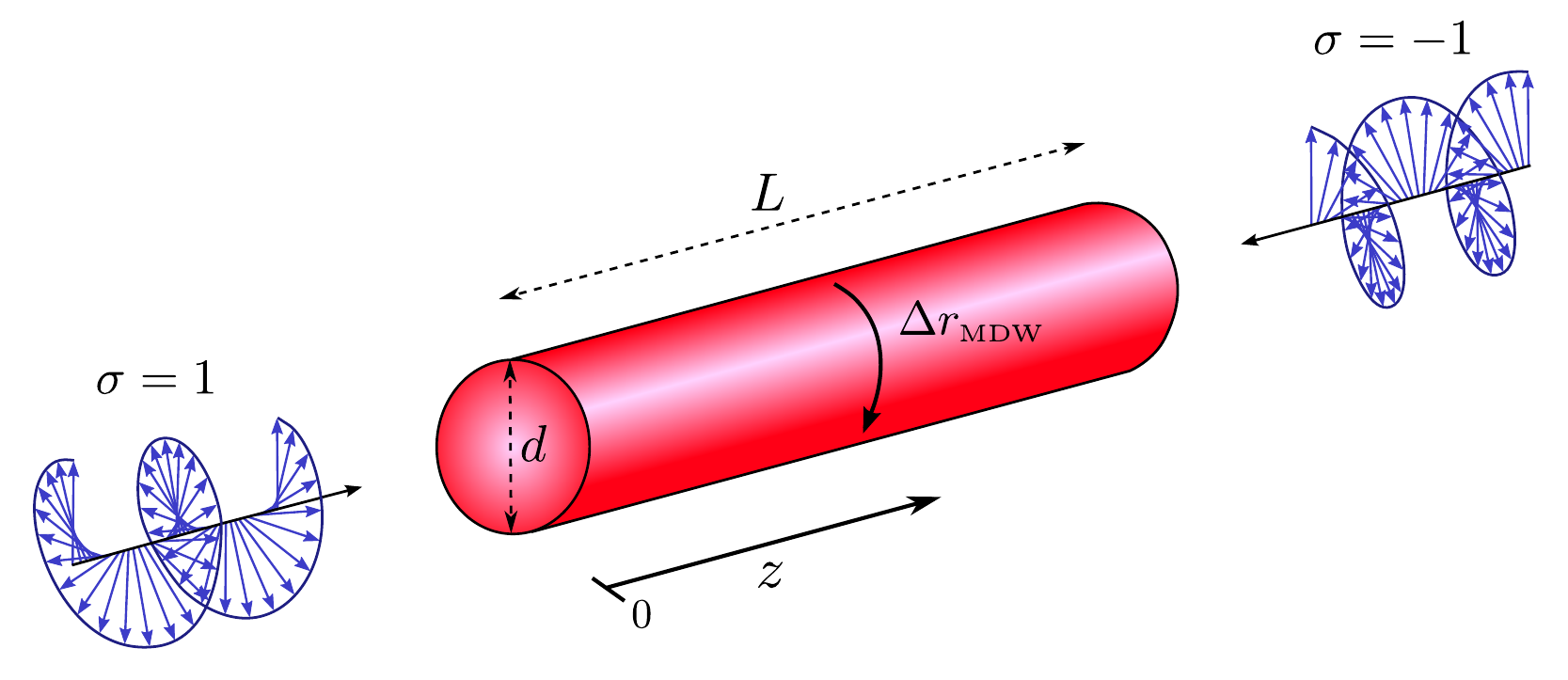}
\caption{\label{fig:setup}
Schematic experimental setup for measuring
the azimuthal rotation displacement $\Delta r_\mathrm{MDW}$ of an optical fiber due to the MDWs
resulting from the optical forces of
circularly polarized continuous wave laser beams.
One of the beams has right-handed polarization ($\sigma=1$)
while the polarization of the other beam is left-handed ($\sigma=-1$).
The rotation is to be measured in the time-scale of $0.01$ s
after the laser beams have been switched on. In longer time
scales, the angular momentum transfer due to the optical
absorption starts to dominate the MDW effect.}
\end{figure}

A schematic illustration of the proposed setup is presented
in Fig.~\ref{fig:setup}. The silicon fiber is expected to rotate
due to the optical forces of two circularly polarized light
beams propagating in opposite directions. One of the beams
has right-handed polarization while the polarization of the other
beam is left-handed. This way the azimuthal atomic displacements of the
MDWs of the two beams add up instead of canceling each other.
The fiber and its ends are allowed to rotate freely
so that the fiber does not experience external torques.
As there are two beams of equal intensities propagating in
opposite directions, the total longitudinal atomic displacement
in the middle of the fiber, studied in Ref.~\cite{Partanen2017e},
is approximately zero. In addition to the atomic displacements due
to the MDWs, also the optical absorption in the fiber contributes to
the total rotation of the fiber. This will be studied below.
On the other hand, some secondary
effects, such as thermal expansion, cannot significantly influence
the rotation of the fiber since they are not azimuthal in nature.

In the following calculations aimed to give order of magnitude estimates,
we use the bulk value of the refractive index.
In more detailed calculations, it should be
replaced with the effective refractive index corresponding
to the waveguide design. The waveguide
dispersion should also be taken into account.
Due to the interface
effects, the fiber diameter cannot be directly
compared with the fiber diameter of our calculations. The
fiber diameter should also be corrected
for the possible cladding layer, metallic coating, and other
factors that influence the spreading of the pulse energy in the
transverse direction. All these factors
can be easily accounted for in the OCD simulations.
Detailed calculations are presented
in a separate work \cite{Partanen2018b}.

Using Eq.~\eqref{eq:Jmdw}, we obtain that the
total angular momentum of the MDW is
$J_\mathrm{MDW}=(1-1/n^2)\sigma U_\mathrm{in}/\omega_0$,
where $U_\mathrm{in}$ is the total electromagnetic energy
inside the fiber. In terms of the time-averaged intensity
$I$ of a single beam in the fiber, one can write
$U_\mathrm{in}=2n\pi R^2LI/c$, where $R$ and $L$ are
the radius and length of the cylindrical fiber, respectively.
On the other hand, the angular momentum
of the MDW can be written as $J_\mathrm{MDW}=I_\mathrm{mi}\Omega$, where
$I_\mathrm{mi}=\frac{1}{2}\pi\rho_0 R^4L$ is the moment
of inertia and $\Omega$ is the angular velocity of the cylindrical fiber.
Setting the two expressions of the angular momentum above as
equal gives the angular velocity of the fiber as
$\Omega=4\sigma (n-1/n) I/(c\,\omega_0\rho_0 R^2)$.
The corresponding azimuthal atomic displacement on the surface
of the fiber as a function of time is then given by
$\Delta r_\mathrm{MDW}=R\Omega t=4\sigma (n-1/n) It/(c\,\omega_0\rho_0 R)$.

Respectively, one can estimate the azimuthal atomic displacement
on the surface of the fiber following from the optical absorption.
The total angular momentum absorbed by the fiber in time $\Delta t$ is given by
$\Delta J_\mathrm{abs}=(1-e^{-\alpha L})(2\sigma\pi R^2I\Delta t/\omega_0)\approx 2\pi\alpha\sigma L R^2I\Delta t/\omega_0$,
where $\alpha$ is the small absorption coefficient of the medium.
On the other hand, we have $\Delta J_\mathrm{abs}=I_\mathrm{mi}\Delta\Omega_\mathrm{abs}$.
Therefore, the angular acceleration related to the optical absorption is given by
$\alpha_\mathrm{abs}=\Delta\Omega_\mathrm{abs}/\Delta t=4\alpha\sigma I/(\omega_0\rho_0 R^2)$.
The corresponding azimuthal atomic displacement on the surface
of the fiber as a function of time is then given by
$\Delta r_\mathrm{abs}=\frac{1}{2}R\alpha_\mathrm{abs} t^2=2\alpha\sigma It^2/(\omega_0\rho_0 R)$.

Above, we have derived two contributions to the longitudinal atomic
displacement on the surface of the fiber. The effect of the MDW depends linearly
on time while the effect of the absorption has a quadratic time dependence.
Therefore, the MDW effect dominates at small time scales while the
effect of the absorption becomes dominant in the course of time.
The time scale at which these effects have equal magnitude
can be solved by setting $\Delta r_\mathrm{MDW}=\Delta r_\mathrm{abs}$, which leads
to $t_\mathrm{eq}=2(n-1/n)/(c\alpha)$.
In the case of silicon, absorption is very low at $\lambda_0=1550$ nm.
The measurements by Schinke \emph{et al.}~\cite{Schinke2015} and Green \cite{Green2008}
for $\lambda_0=1450$ nm give $\alpha\approx10^{-8}$ cm$^{-1}$ and the absorption
is known to decrease towards $\lambda_0=1550$ nm. Therefore, we can conservatively estimate
$\alpha=10^{-8}$ cm$^{-1}$. This gives $t_\mathrm{eq}=21$ ms.

As shown above, the atomic displacement on the surface of the fiber
depends linearly on the field intensity.
One must note that the field intensity in the experiment
cannot be arbitrarily high. For silicon with $\lambda_0=1550$ nm,
the bulk value of the breakdown threshold energy
density has been reported to be $u_\mathrm{th}=13.3$ J/cm$^3$ \cite{Cowan2006},
which corresponds to the threshold irradiance of
$I_\mathrm{th}=u_\mathrm{th}c/n=1.1\times 10^{11}$ W/cm$^2$.
Using the fiber of diameter $d=2.5$ $\mu$m, assuming
$I=\frac{1}{100}I_\mathrm{th}$, and setting $t=t_\mathrm{eq}$,
we obtain $\Delta r_\mathrm{MDW}=\Delta r_\mathrm{abs}=2.8$ nm,
which corresponds to the total atomic displacement of $5.6$ nm.
The atomic displacement of this magnitude should be
feasible to measure. As the time dependencies of
the MDW and absorption contributions are different,
one should also be able to distinguish the magnitudes
of both effects by observing the change of the atomic
displacement as a function of time.

\begin{figure}
\includegraphics[width=\columnwidth]{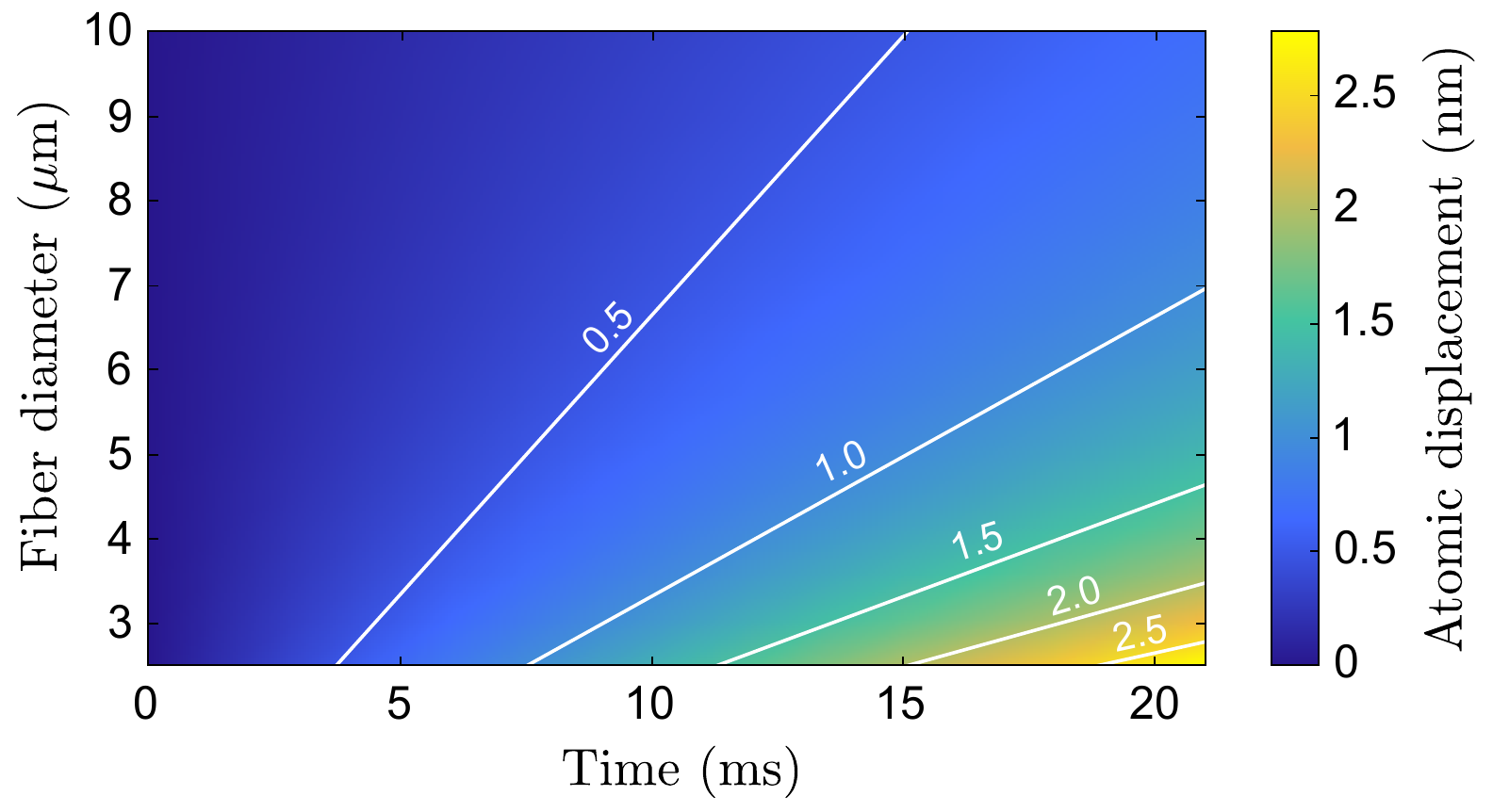}
\vspace{-0.5cm}
\caption{\label{fig:rotationdisplacement}
The calculated total azimuthal atomic displacement due to the MDW
on the surface of a cylindrical silicon fiber as a function of time and the
fiber diameter. The assumed field intensity is $I=1.1\times 10^9$ W/cm$^2$.
The refractive index of silicon for the assumed vacuum wavelength of $\lambda_0=1550$ nm
is $n=3.48$. In the time range shown in the figure, the atomic displacement
due to the MDW dominates the atomic displacement due to the optical
absorption.}
\vspace{-0.1cm}
\end{figure}

Figure \ref{fig:rotationdisplacement} shows the calculated
azimuthal atomic displacement $\Delta r_\mathrm{MDW}$
on the fiber surface due to the MDW as a function of time and the fiber diameter.
As reasoned above, the atomic displacement is largest for small fiber diameters
and it has a linear time dependence.
In the time range $[0,t_\mathrm{eq}]$ shown in the figure, the atomic displacement
$\Delta r_\mathrm{MDW}$ due to the MDW dominates the atomic displacement
$\Delta r_\mathrm{abs}$ due to the optical
absorption. The contribution $\Delta r_\mathrm{abs}$ is not shown
in the figure. The simulated nanometer-scale atomic displacements
strongly support the experimental feasibility of
the measurement of the azimuthal atomic displacement of the MDW.

It is also important to note that
the relatively high intensity, which is needed to produce measurable atomic
displacements, can lead to nonlinear effects that must also be
accounted for in the implementation of the experiment.
To reduce the intensity of single frequency components,
one could well use a broad spectrum
only requiring that the absorption coefficient is small
for the frequencies used. This would allow reducing nonlinear
effects, such as the stimulated Brillouin scattering,
which is probably the first nonlinear effects that turns on.

\section{\label{sec:conclusions}Conclusions}

In conclusion, we have used the MP theory of light to analyze
how the angular momentum of light is shared between the field and matter.
The optical force density,
which gives rise to the momentum of the MDW \cite{Partanen2017c,Partanen2017e},
also describes the optical torque. Consequently, in many
photonic materials, most of the total angular momentum of light,
both OAM and SAM, is not
carried by the electromagnetic field but by the MDW. For
the electromagnetic field, one obtains a share of the angular momentum that
is generally a fraction of $\hbar$.
In contrast, the total angular momentum of the MP, where
the field is coupled to the MDW, is an integer multiple of $\hbar$.
This suggests that the angular momentum of a single quantum in a nondispersive medium
has coupled field
and medium components. The same coupling is found for OAM and SAM.
The coupling that makes it impossible to measure
separately the single quantum angular momentum components
should not be confused with the well-known quantum entanglement of
eigenstates, but it is sooner related to the covariance principle of the special
theory of relativity, which prevents a bare photon from
existing in a nondispersive medium \cite{Partanen2017c,Partanen2017e}.
The generalization of the conventional quantum optical field theory description
of light for the presentation of the MDW coupling,
studied classically in the present work, provides an interesting
topic for further work.

Our results strongly suggest that a single quantum of light in a nondispersive medium
should be understood to include coupled field and medium components.
Thus, a single quantum becomes Lorentz invariant
and fulfills the constant center of energy law of an isolated system.
Compared to earlier theoretical models neglecting the atomic MDW,
the physical picture of the angular momentum of light emerging from our theory
is fundamentally more general.
In previous works, the correct total angular
momentum of light in a medium has been obtained only
by assuming that the pure electromagnetic field or the electronic
polariton state carries the Minkowski momentum.
The results of the MP theory of light show that this assumption
is not justified.
We have also pointed out that the
contribution of the atomic MDW to the total angular momentum
of light is experimentally verifiable.
In this work, we have assumed a nondispersive medium, but the results
can be straightforwardly generalized for dispersive media
as done in the case of linear momentum in Ref.~\cite{Partanen2017e}.

One interesting question that remains as a topic of further work
is whether slow-light media could be used to amplify
the MDW effect to allow its experimental verification
in the same way as slow-light media have been used to 
amplify the rotary photon drag \cite{FrankeArnold2011} and
Fresnel drag effects \cite{Safari2016}. This is, however,
not obvious as the effects are fundamentally different.
In the rotary photon drag and Fresnel drag experiments,
light propagates through materials that are set in
notable motion by external forces and the effects observed follow from
this movement. In contrast, in the case of the MDW effect,
the medium in which light propagates is initially
at rest and the medium atoms become very slightly displaced
by the optical force of the light field itself.

Very accurate technologies are presently becoming
available for measuring the small atomic displacements
on the surfaces of solid media. For example,
the picometer-size atomic displacements of
elastic waves generated at a highly-reflective mirror
interface have been a subject of a very recent careful study \cite{Pozar2018}.
Therefore, we expect that by using a suitable setup,
these technologies might also allow probing the atomic
displacements generated by light while it is propagating inside the medium.

\begin{acknowledgments}
This work has been funded in part by the Academy of Finland under Contracts No.~287074
and No.~318197 and the Aalto Energy Efficiency Research Programme.
We also want to thank Toma\ifmmode\check{z}\else\v{z}\fi{} Po\ifmmode\check{z}\else\v{z}\fi{}ar
and Nelson G. C. Astrath for the recent communication, where we became aware of
the old paper of Poynting \cite{Poynting1911}, the discussion of which has
been added in the present work.
\end{acknowledgments}

\appendix

\section{\label{apx:separations}Angular momentum separations}

\subsection{Spin and orbital angular momenta}

In a nondispersive dielectric medium, the total angular momentum
of the electromagnetic field in Eq.~\eqref{eq:Jfield}
can be written in terms of the vector potential $\mathbf{A}$ as \cite{Andrews2013,Piccirillo2013,vanEnk1992,vanEnk1994}
\begin{align}
 \mathbf{J}_\mathrm{field}
 &=\varepsilon_0\int E_i(\mathbf{r}\times\nabla)A_id^3r
 +\varepsilon_0\int\mathbf{E}\times\mathbf{A}d^3r\nonumber\\
 &\hspace{0.5cm}-\varepsilon_0\int(\mathbf{r}\times\mathbf{A})\mathbf{E}\cdot d\mathbf{S},
\label{eq:Jfield2}
\end{align}
where $\varepsilon_0$ is the permittivity of vacuum. The vector potential
$\mathbf{A}$ is related to the electric and magnetic fields by
the conventional relations $\mathbf{E}=-\partial\mathbf{A}/\partial t$
and $\mathbf{B}=\nabla\times\mathbf{A}$ with $\mathbf{B}=\mu_0\mathbf{H}$,
where $\mu_0$ is the permeability of vacuum.
In the first term on the right hand side of Eq.~\eqref{eq:Jfield2},
we have used the Einstein summation convention.

For a light pulse, the fields approach zero at infinity. Therefore,
the surface integral in the last term of Eq.~\eqref{eq:Jfield2} will be zero,
and we can identify the OAM and SAM terms as \cite{Andrews2013,Piccirillo2013,Nieminen2008}
\begin{equation}
 \mathbf{L}_\mathrm{field}=\varepsilon_0\int E_i(\mathbf{r}\times\nabla) A_id^3r,
 \label{eq:Lfield}
\end{equation}
\begin{equation}
 \mathbf{S}_\mathrm{field}=\varepsilon_0\int\mathbf{E}\times\mathbf{A}d^3r.
 \label{eq:Sfield}
\end{equation}
One can observe that these definitions of OAM and SAM in terms of the vector
potential are in general gauge dependent. However, for fields that are exactly transverse,
gauge invariance is obtained if $\mathbf{A}$ is defined to
be the gauge invariant transverse vector potential \cite{Piccirillo2013,CohenTannoudji1989}.

The transverse field approximation is in many cases well justified.
In the commonly used paraxial approximation of light beams, the electric and magnetic fields also
have longitudinal components \cite{Carnicer2012,Lax1975,Allen2011}.
The magnitudes of these
components are, however, typically negligible in comparison with the transverse field
components. Therefore, in the paraxial approximation, Eqs.~\eqref{eq:Lfield} and \eqref{eq:Sfield}
can be used to separate the total angular momentum of light
into the OAM and SAM parts reasonably accurately, but not exactly.
In general, this separation is gauge dependent and only
the total angular momentum of
the electromagnetic field in Eq.~\eqref{eq:Jfield} has a well-defined physical meaning.
In this work, we use the general definition of electromagnetic angular momentum in Eq.~\eqref{eq:Jfield}
in all our calculations. We also show that the movement of atoms
in the MDW driven by the field carries a substantial part
of the total angular momentum of light in a medium
as described by Eq.~\eqref{eq:Jmdw}.

\subsection{External and internal angular momenta}

In the case of a localized light pulse whose center of energy
has a position vector $\mathbf{r}_0$, by a change of variables
$\mathbf{r}\rightarrow\mathbf{r}_0+\mathbf{r}'$, we can write
Eq.~\eqref{eq:Jfield} as
\begin{equation}
 \mathbf{J}_\mathrm{field}=\mathbf{J}_\mathrm{field,ext}+\mathbf{J}_\mathrm{field,int},
\end{equation}
where $\mathbf{J}_\mathrm{field,ext}$ and
$\mathbf{J}_\mathrm{field,int}$ are, respectively, the external
and internal angular momenta of the field \cite{Lenstra1982,Jauregui2005,Calvo2006,LiCF2009}.
The external angular momentum is contributed by the OAM only,
while the internal angular momentum is contributed by both the OAM and SAM.
The origin-dependent external angular momentum is given by
$\mathbf{J}_\mathrm{field,ext}=\mathbf{r}_0\times\mathbf{P}_\mathrm{field}$,
where $\mathbf{P}_\mathrm{field}=\int\mathbf{g}_\mathrm{field}d^3r$ is the
total linear momentum of the electromagnetic field.
Therefore, the definition of the external angular momentum
corresponds to the classical definition of the angular momentum of a point particle.
It gives zero if the angular momentum is calculated with
respect to the propagation axis in which case
$\mathbf{r}_0$ is parallel to $\mathbf{P}_\mathrm{field}$.
The internal angular momentum is given by
$\mathbf{J}_\mathrm{field,int}=\int\mathbf{r}'\times\mathbf{g}_\mathrm{field}d^3r'$
and it does not depend on the choice of the origin.
In the case of structured fields, the internal angular
momentum $\mathbf{J}_\mathrm{field,int}$ can
be nonzero \cite{Yao2011}.
This is also shown in the simulation results.
In the simulations, we make the conventional choice that the origin
lies in the optical axis so that the external angular momentum is
zero. Therefore, we directly study the internal angular momentum
quantities.

\section{\label{apx:approximation}Monochromatic field approximation of light pulses}

\subsection{Linear polarization}

In the monocromatic field limit with $\Delta k_0\ll k_0$,
we can approximate the
$k$-dependent parts of the integrands in Eqs.~\eqref{eq:efield1} and \eqref{eq:hfield1}
apart from the last part $u(k)e^{i[kz-\omega(k) t]}$
with the central frequency values as $\omega(k)\approx\omega_0$
and $k\approx k_\mathrm{0,med}=nk_0$. After this, the integrals
over $k$ in Eqs.~\eqref{eq:efield1} and \eqref{eq:hfield1} can
be evaluated analytically leading to the fields given by
\begin{align}
 \mathbf{E}_{p,l}(\mathbf{r},t)
&\approx \mathrm{Re}\Big[i\omega_0\Big(u_{p,l}\hat{\mathbf{x}}+\frac{i\partial u_{p,l}}{nk_0\partial x}\hat{\mathbf{z}}\Big)e^{i(nk_0z-\omega_0t)}\Big]\nonumber\\
&\hspace{0.4cm}\times e^{-(n\Delta k_0)^2(z-ct/n)^2/2},
\end{align}
\begin{align}
 \mathbf{H}_{p,l}(\mathbf{r},t)
&\approx \mathrm{Re}\Big[\frac{ink_0}{\mu_0}\Big(u_{p,l}\hat{\mathbf{y}}+\frac{i\partial u_{p,l}}{nk_0\partial y}\hat{\mathbf{z}}\Big)e^{i(nk_0z-\omega_0t)}\Big]\nonumber\\
&\hspace{0.4cm}\times e^{-(n\Delta k_0)^2(z-ct/n)^2/2}.
\end{align}
In the calculation of the optical force density by using Eq.~\eqref{eq:opticalforcedensity},
we need the Poynting vector. We perform part of our simulations by using
the actual instantaneous Poynting vector and part of the simulations by using its time average
over the harmonic cycle.
As shown in Ref.~\cite{Partanen2017c}, we can approximate the instantaneous Poynting
vector of a light pulse with its time-average over the harmonic cycle
without losing accuracy in the calculation of the total linear and angular momenta,
the transferred mass, and the averaged atomic displacements and velocities.
This time-averaged Poynting vector is given by
\begin{align}
 &\langle\mathbf{E}_{p,l}(\mathbf{r},t)\times\mathbf{H}_{p,l}(\mathbf{r},t)\rangle\nonumber\\
&\approx\frac{n\omega_0^2}{2\mu_0c}\mathrm{Re}\Big(\frac{iu_{p,l}\partial u_{p,l}^*}{nk_0\partial x}\hat{\mathbf{x}}-\frac{iu_{p,l}^*\partial u_{p,l}}{nk_0\partial y}\hat{\mathbf{y}}+|u_{p,l}|^2\hat{\mathbf{z}}\Big)\nonumber\\
&\hspace{0.4cm}\times e^{-(n\Delta k_0)^2(z-ct/n)^2}\nonumber\\
&=-\frac{i\omega_0}{4\mu_0}\Big(u_{p,l,k}^*\nabla u_{p,l,k}-u_{p,l,k}\nabla u_{p,l,k}^*\Big)e^{-(n\Delta k_0)^2(z-ct/n)^2}.
\end{align}
Here, the first expression on the right is the explicit representation in the Cartesian basis
and the second appealing expression is obtained by using $u_{p,l,k}=u_{p,l}e^{ink_0z}$.
For a LG beam without Gaussian form in the longitudinal direction,
the corresponding result has been presented in the case of vacuum in Ref.~\cite{Allen1992b}.

\subsection{Circular polarization}

Using the same approximation as explained in the case of linear polarization above,
the electric and magnetic fields of a right circularly polarized $\mathrm{LG}_{pl}$ pulse
in Eqs.~\eqref{eq:efield2} and \eqref{eq:hfield2} can be written as
\begin{align}
 \mathbf{E}_{p,l}(\mathbf{r},t)
&\approx \frac{1}{\sqrt{2}}\mathrm{Re}\Big[i\omega_0\Big(u_{p,l}\hat{\mathbf{x}}+\frac{i\partial u_{p,l}}{nk_0\partial x}\hat{\mathbf{z}}\Big)e^{i(nk_0z-\omega_0t)}\nonumber\\
&\hspace{0.4cm}+i\omega_0\Big(u_{p,l}\hat{\mathbf{y}}+\frac{i\partial u_{p,l}}{nk_0\partial y}\hat{\mathbf{z}}\Big)e^{i(nk_0z-\omega_0t+\pi/2)}\Big]\nonumber\\
&\hspace{0.4cm}\times e^{-(n\Delta k_0)^2(z-ct/n)^2/2},
\end{align}
\begin{align}
 \mathbf{H}_{p,l}(\mathbf{r},t)
&\approx \frac{1}{\sqrt{2}}\mathrm{Re}\Big[\frac{ink_0}{\mu_0}\Big(u_{p,l}\hat{\mathbf{y}}+\frac{i\partial u_{p,l}}{nk_0\partial y}\hat{\mathbf{z}}\Big)e^{i(nk_0z-\omega_0t)}\nonumber\\
&\hspace{0.4cm}-\frac{ink_0}{\mu_0}\Big(u_{p,l}\hat{\mathbf{x}}+\frac{i\partial u_{p,l}}{nk_0\partial x}\hat{\mathbf{z}}\Big)e^{i(nk_0z-\omega_0t+\pi/2)}\Big]\nonumber\\
&\hspace{0.4cm}\times e^{-(n\Delta k_0)^2(z-ct/n)^2/2}.
\end{align}

The Poynting vector time-averaged over the harmonic cycle is then given by
\begin{align}
 &\langle\mathbf{E}_{p,l}(\mathbf{r},t)\times\mathbf{H}_{p,l}(\mathbf{r},t)\rangle\nonumber\\
&\approx\frac{n\omega_0^2}{2\mu_0c}\mathrm{Re}\Big[\frac{u_{p,l}}{nk_0}\Big(\frac{\partial u_{p,l}^*}{\partial y}+i\frac{\partial u_{p,l}^*}{\partial x}\Big)\hat{\mathbf{x}}\nonumber\\
&\hspace{0.4cm}-\frac{u_{p,l}^*}{nk_0}\Big(\frac{\partial u_{p,l}}{\partial x}+i\frac{\partial u_{p,l}}{\partial y}\Big)\hat{\mathbf{y}}+|u_{p,l}|^2\hat{\mathbf{z}}\Big]\nonumber\\
&\hspace{0.4cm}\times e^{-(n\Delta k_0)^2(z-ct/n)^2}\nonumber\\
&=-\frac{i\omega_0}{4\mu_0}\Big(u_{p,l,k}^*\nabla u_{p,l,k}-u_{p,l,k}\nabla u_{p,l,k}^*-i\frac{\partial |u_{p,l,k}|^2}{\partial r}\hat{\boldsymbol{\phi}}\Big)\nonumber\\
&\hspace{0.4cm}\times e^{-(n\Delta k_0)^2(z-ct/n)^2},
\end{align}
where $\hat{\boldsymbol{\phi}}=-\sin(\phi)\hat{\mathbf{x}}+\cos(\phi)\hat{\mathbf{y}}$ is the azimuthal unit vector.
Again, the first expression on the right is the explicit representation in the Cartesian basis
and the second appealing expression is obtained by using $u_{p,l,k}=u_{p,l}e^{ink_0z}$.
In the last expression, the first two terms inside the parentheses are polarization independent
and relate to the OAM while the last term is polarization dependent and relates to the SAM.
For a LG beam without Gaussian form in the longitudinal direction,
the corresponding result has been presented in the case of vacuum in Ref.~\cite{Allen1992b}.

\end{document}